\documentclass[12pt]{article}
\usepackage{graphicx}
\usepackage{epsfig}
\def\bvv{B \to V_1 V_2}
\def\barpk{{\raise.35ex\hbox
{${\sss (}$}}--{\raise.35ex\hbox{${\sss )}$}}}
\def\bbarp{\hbox{$B$\kern-0.9em\raise1.4ex\hbox{\barpk}}}
\def\beq{\begin{equation}}
\def\eeq{\end{equation}}
\def\bea{\begin{eqnarray}}
\def\eea{\end{eqnarray}}
\def\nn{\nonumber}
\def\sss{\scriptscriptstyle}
\def\roughly#1{\mathrel{\raise.3ex\hbox
{$#1$\kern-.75em\lower1ex\hbox{$\sim$}}}}

\def\bd{B_d^0}
\def\bs{B_s}
\def\bdbar{{\bar B}_d^0}
\def\btod{{\bar b} \to {\bar d}}
\def\btos{{\bar b} \to {\bar s}}

\def\fT{f_{\sss T}}
\def\fL{f_{\sss L}}
\def\fTfL{f_{\sss T}/f_{\sss L}}
\def \Kbar{\bar K}

\def\Abar{{\bar A}}

\def\epjc#1#2#3{{ Eur.\ Phys.\ J.}\ {\bf C#1}, #3 (#2)}

\def\plb#1#2#3{{ Phys.\ Lett.} {\bf #1B}, #3 (#2)}

\def\newprd#1#2#3{{ Phys.\ Rev.} {\bf D#1}, #3 (#2)}
\def\prl#1#2#3{{ Phys.\ Rev.\ Lett.} {\bf #1}, #3 (#2)}

\begin{document}

\begin{flushright}
UMiss-HEP-2007-03 \\
UdeM-GPP-TH-07-160 \\
\end{flushright}

\begin{center}
\bigskip
{\Large \bf \boldmath Testing Explanations of the $B\to \phi K^*$
Polarization Puzzle} \\
\bigskip
\bigskip
{\large
Alakabha Datta $^{a,}$\footnote{datta@phy.olemiss.edu},
Andrei V. Gritsan $^{b,}$\footnote{gritsan@jhu.edu},
David London $^{c,}$\footnote{london@lps.umontreal.ca}, \\
Makiko Nagashima $^{c,}$\footnote{makiko@lps.umontreal.ca},
and Alejandro Szynkman $^{c,}$\footnote{szynkman@lps.umontreal.ca}
}
\end{center}

\begin{flushleft}
~~~~~~~~~~~$a$: {\it Dept.\ of Physics and Astronomy, 108 Lewis Hall,}\\
~~~~~~~~~~~~~~~{\it University of Mississippi, Oxford, MS 38677-1848, USA}\\
~~~~~~~~~~~$b$: {\it Dept.\ of Physics and Astronomy, Johns Hopkins
University,}\\
~~~~~~~~~~~~~~~{\it Baltimore, MD 21218, USA}\\
~~~~~~~~~~~$c$: {\it Physique des Particules, Universit\'e
de Montr\'eal,}\\
~~~~~~~~~~~~~~~{\it C.P. 6128, succ. centre-ville, Montr\'eal, QC,
Canada H3C 3J7}\\
\end{flushleft}

\begin{center}
\bigskip (\today)
\vskip0.5cm {\Large Abstract\\} \vskip3truemm
\parbox[t]{\textwidth}{$B\to\phi K^*$ ($\btos$) is three separate
decays, one for each polarization of the final-state vector mesons
(one longitudinal, two transverse). It is observed that the fraction
of transverse decays, $\fT$, and the fraction of longitudinal decays,
$\fL$, are roughly equal: $\fTfL \simeq 1$, in opposition to the naive
expectation that $\fT \ll \fL$. If one requires a single explanation
of all polarization puzzles, two possibilities remain within the
standard model: penguin annihilation and rescattering. In this paper
we examine the predictions of these two explanations for $\fTfL$ in
$\btod$ decays. In $B \to \rho\rho$ decays, only $\bd \to
\rho^0\rho^0$ can possibly exhibit a large $\fTfL$. In $B$ decays
related by U-spin, we find two promising possibilities: (i) $B^+ \to
K^{*0} \rho^+$ ($\btos$) and $B^+ \to \Kbar^{*0} K^{*+}$ ($\btod$) and
(ii) $\bs \to K^{*0} \Kbar^{*0}$ ($\btos$) and $\bd \to \Kbar^{*0}
K^{*0}$ ($\btod$). The measurement of $\fTfL$ in these pairs of decays
will allow us to test penguin annihilation and rescattering. Finally,
it is possible to distinguish penguin annihilation from rescattering
by performing a time-dependent angular analysis of $\bd \to \Kbar^{*0}
K^{*0}$.}
\end{center}


\thispagestyle{empty}
\newpage
\setcounter{page}{1}
\baselineskip=14pt

\section{Introduction}

The $B$-factories BABAR and Belle, along with Tevatron experiments,
have been operating for several years now, making many measurements of
various $B$ decays. As always, the hope is to find results which are
in contradiction with the expectations of the standard model (SM) and
which therefore show evidence for the presence of physics beyond the
SM. To date, there have been several hints of such new physics in
$\btos$ transitions, though none has been statistically significant.

One intriguing puzzle was first seen in $B\to\phi K^*$ decays
\cite{phiK*}.  In this decay the final-state particles are vector
mesons. Thus, when the spin of the vector mesons is taken into
account, this decay is in fact three separate decays, one for each
polarization (one longitudinal, two transverse). Naively, the
transverse amplitudes are suppressed by a factor of size $m_{\sss V}/m_{\sss
B}$ ($V$ is one of the vector mesons) with respect to the longitudinal
amplitude. As such, one expects the fraction of transverse decays,
$\fT$, to be much less than the fraction of longitudinal decays,
$\fL$. However, it is observed that these two fractions are roughly
equal: $\fTfL (B\to\phi K^{*}) \simeq 1$ (see Table~\ref{table1}).

\begin{table}[tbh]
\center
{
\small
\begin{tabular}{lccccc}
\hline
\vspace{-0.3cm}\\
Mode & ${\cal B}(10^{-6})$ & $f_{\sss L}$ & $f_\perp$ & $\phi_\parallel-\pi$  &
$\phi_\perp-\pi$  \\
\vspace{-0.3cm}\\
\hline\hline
\vspace{-0.3cm}\\
$\phi K^{*0}$ \cite{phiK*0, belle:phikst, cdf:phikst} &
  ${ 9.5}\pm{0.9}$ & ${ 0.49}\pm{0.04}$ & ${ 0.27}^{+0.04}_{-0.03}$ &
${-0.73}^{+0.18}_{-0.16}$ & ${-0.62}\pm0.17$ \\
$\phi K^{*+}$ \cite{phiK*, belle:phikst, babar:phikstpl} &
  ${ 10.0}\pm 1.1$ & ${ 0.50}\pm 0.05$ & ${0.20}\pm 0.05$ & 
$-0.80\pm0.17$ & $-0.56\pm0.17$ \\
$\rho^+ K^{*0}$ \cite{belle:rhokst, babar:rhokst} &
${ 9.2}\pm 1.5$ &  ${ 0.48}\pm 0.08$ & --& --& --\\
$\rho^0 K^{*0}$ \cite{babar:rhokst} &
${ 5.6}\pm{1.6}$  & ${ 0.57}\pm0.12$  & --& --& --\\
$\rho^- K^{*+}$ \cite{babar:rhokst} &
{$<{ 12.0}$} & -- & --& --& --\\
$\rho^0 K^{*+}$ \cite{babar:rhokst} &
({$3.6^{+1.9}_{-1.8}$}) & $(0.9\pm 0.2)$ & --& --& --\\
$\omega K^{*0}$ \cite{omegakst} &
($2.4\pm1.3$) & $(0.71^{+0.27}_{-0.24})$ & --& --& --\\
$\omega K^{*+}$ \cite{omegakst}  &
$<{ 3.4}$ & -- & --& --& --\\
\vspace{-0.3cm}\\
\hline\hline
\end{tabular}
}
\caption{Measurements of the branching fraction ${\cal B}$,
longitudinal polarization fraction $\fL$, fraction of parity-odd
transverse amplitude $f_\perp$, and phases of the two transverse
amplitudes $\phi_\parallel$ and $\phi_\perp$ (rad) with respect to the
longitudinal amplitude, for $B\to\phi K^*$, $\rho K^*$, and $\omega
K^*$, expected to proceed through a $\btos$ transition \cite{pdg,
hfag}.  Numbers in parentheses indicate observables measured with less
than 4$\sigma$ significance.  We quote the solution of
$\phi_\parallel$ and $\phi_\perp$ according to the phase ambiguity
resolved by BABAR \cite{phiK*0,babar:phikstpl}.  For a complete list
of up to 12 parameters measured, including CP-violating observables,
see references quoted.  }
\label{table1}
\end{table}

Within the SM, there are three potential explanations of the observed
$\fTfL$ ratio, described in more detail in Sec.~2: penguin
annihilation \cite{Kagan}, rescattering \cite{soni, SCET}, and enhanced
penguin contributions due to the dipole operator
\cite{Beneke,HouMakiko}. Assuming only one of these explanations is
valid, enhanced dipole-operator contributions are ruled out by the
observed large $\fTfL$ in $B^+ \to \rho^+ K^{*0}$ (see
Table~\ref{table1}). However, the other two are in agreement with all
observed data. In this paper, we explore ways of testing these
explanations.

These two explanations account for a large $\fTfL$ in $\btos$
decays. However, the key point is that a large $\fTfL$ is also
predicted in certain $\btod$ decays \cite{DMV}. We examine these
predictions. The measurement of $\fTfL$ in these $\btod$ decays will
allow us to test penguin annihilation and rescattering as the
explanations of the observed $\fTfL$ ratio in $B\to\phi K^*$ decays,
or maybe even rule them out.

In Sec.~2, we describe in more detail the SM explanations of the
observed $\fTfL$ ratio, along with constraints from present
data. Sec.~3 contains the predictions for $\fTfL$ in $B \to \rho\rho$
decays. Only $\bd \to \rho^0\rho^0$ is expected to possibly exhibit a
large $\fTfL$. In Sec.~4, we examine $\fTfL$ for various $B$ decays
related by U-spin. Sec.~5 contains a method for distinguishing penguin
annihilation from rescattering. We conclude in Sec.~6.

\section{\boldmath Explanations of $\fTfL$ in $B\to\phi K^*$}

We focus here on $B\to V_1V_2$ decays ($V_i$ is a vector meson). This
is really three decays, one for each polarization of the final state.
Here it is useful to use the linear polarization basis, where one
decomposes the decay amplitude into components in which the
polarizations of the final-state vector mesons ($\varepsilon_i^*$) are
either longitudinal ($A_0$), or transverse to their directions of
motion and parallel ($A_\|$) or perpendicular ($A_\perp$) to one
another. The amplitude for this decay is given by \cite{DDLR,CW}
\beq
M = A_0 \varepsilon_1^{*\sss L} \cdot \varepsilon_2^{*\sss L} 
- {1 \over \sqrt{2}} A_\| {\vec\varepsilon}_1^{*\sss T} \cdot
  {\vec\varepsilon}_2^{*\sss T}
- {i \over \sqrt{2}} A_\perp {\vec\varepsilon}_1^{*\sss T} \times
  {\vec\varepsilon}_2^{*\sss T} \cdot {\hat p} ~,
\eeq
where ${\hat p}$ is the unit vector along the direction of motion of
$V_2$ in the rest frame of $V_1$, $\varepsilon_i^{*\sss L} =
{\vec\varepsilon}_i^* \cdot {\hat p}$, and ${\vec\varepsilon}_i^{*\sss
T} = {\vec\varepsilon}_i^* - \varepsilon_i^{*\sss L} {\hat p}$.  In
this paper we will often use the basis $A_\pm$ for the transverse
polarizations, where $A_\pm=(A_\| \pm A_\perp)/\sqrt{2}$.  Note that,
due to the factor of `$i$' in the amplitude above, $A_+$ and $A_-$
change roles in the CP-conjugate decay ${\bar B}\to {\bar V}_1{\bar
V}_2$: $A_+ \to {\bar A}_-$ and $A_- \to {\bar A}_+$.

The fraction of various types of decay is given by
\beq
\fL = {|A_0|^2 \over |A_0|^2 + |A_+|^2 + |A_-|^2}
~~,~~~~ \fT = {|A_+|^2 + |A_-|^2 \over |A_0|^2 + |A_+|^2 + |A_-|^2} ~,
\eeq
where $f_{\sss T}=(1-f_{\sss L})$,
\beq
f_\perp = {|A_\perp|^2 \over |A_0|^2 + |A_+|^2 + |A_-|^2}  ~~,~~~~
f_\parallel = {|A_\parallel|^2 \over |A_0|^2 + |A_+|^2 + |A_-|^2}  ~,
\eeq
where $f_\parallel=(1-f_{\sss L}-f_\perp)$, and the relative phases are
\beq
\phi_\perp = {\rm arg}(A_\perp/A_0)~~,~~~~
\phi_\parallel = {\rm arg}(A_\parallel/A_0)~.
\eeq
We note that when $\phi_\perp=\phi_\parallel$ and
$f_\perp=f_\parallel$, we have $A_-=0$, which is close to experimental
observation for $B\to\phi K^*$ in Table~\ref{table1}.

In the introduction we noted that there are three SM explanations of
the observed $\fTfL$ in $B\to\phi K^*$ decays. We discuss them in more
detail here.

We begin with penguin annihilation \cite{Kagan}, as shown in
Fig.~1. $B\to\phi K^*$ receives penguin contributions, ${\bar b} {\cal
O} s {\bar q} {\cal O} q$, where $q=u,d$ (${\cal O}$ are Lorentz
structures, and color indices are suppressed). Applying a Fierz
transformation, these operators can be written as ${\bar b} {\cal O}'
q {\bar q} {\cal O}' s $. A gluon can now be emitted from one of the
quarks in the operators which can then produce a pair of $s, \bar{s}$
quarks. These then combine with the ${\bar s}, q$ quarks to form the
final states $\phi K^{*+}~(q=u)$ or $\phi K^{*0}~(q=d)$. 

\vskip-0.3truein
\begin{figure}[htbp]
\begin{center}
\begin{tabular}{cc}
\scalebox{0.8}{\epsfig{file=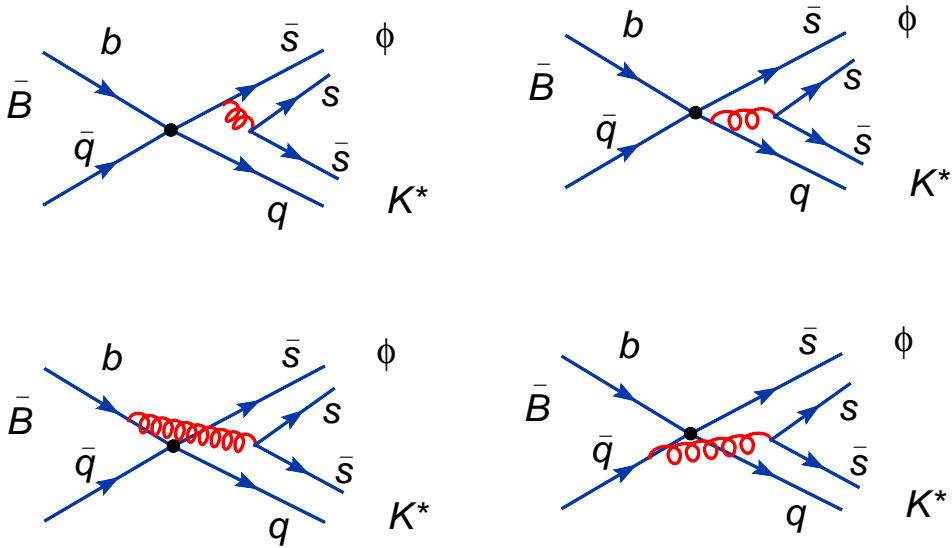}}
\end{tabular}
\caption{The penguin annihilation diagrams.} 
\label{fig:narrow1}
\end{center}
\end{figure}

Normally all annihilation contributions are expected to be small as
they are higher order in the $1/m_b$ expansion, and thus ignored.
However, within QCD factorization (QCDf) \cite{BBNS}, it is plausible
that the coefficients of these terms are large \cite{Kagan}. In QCDf
penguin annihilation is not calculable because of divergences which are
parameterized in terms of unknown quantities. One may choose these
parameters to fit the polarization data in $B\to\phi K^*$ decays.
(Within perturbative QCD \cite{pQCD}, the penguin annihilation is
calculable and can be large, though it is not large enough to explain
the polarization data in $B\to\phi K^*$ \cite{LM}.) Note that the
penguin annihilation term arises only from penguin operators with an
internal $t$ quark.

We now turn to rescattering \cite{soni, SCET}, shown in Fig.~2. It has
been suggested that rescattering effects involving charm intermediate
states, generated by the operator ${\bar b} {\cal O}' c {\bar c} {\cal
O}' s$, can produce large transverse polarization in $B\to\phi K^*$.
A particular realization of this scenario is the following
\cite{soni}.  Consider the decay $B^+ \to D_s^{*+} {\bar D}^{*0}$
generated by the operator ${\bar b} {\cal O}' c {\bar c} {\cal O}' s
$. Since the final-state vector mesons are heavy, the transverse
polarization can be large.  The state $D_s^{*+} {\bar D}^{*0}$ can now
rescatter to $\phi K^{*+}$. If the transverse polarization $T$ is not
reduced in the scattering process, this will lead to $B^+\to\phi
K^{*+}$ with large $\fTfL$. (A similar rescattering effect can take
place for $\bd\to\phi K^{*0}$.)

In principle, rescattering can also take place if $u{\bar u}$ quark
pairs are involved. However, this does not contribute significantly to
$T$. One way to see this is to realize that most intermediate states
are light, so that the transverse polarization is small. Thus, one
cannot obtain a large $\fTfL$ in this case.

\begin{figure}[htbp]
\begin{center}
\begin{tabular}{cc}
\scalebox{0.7}{\epsfig{file=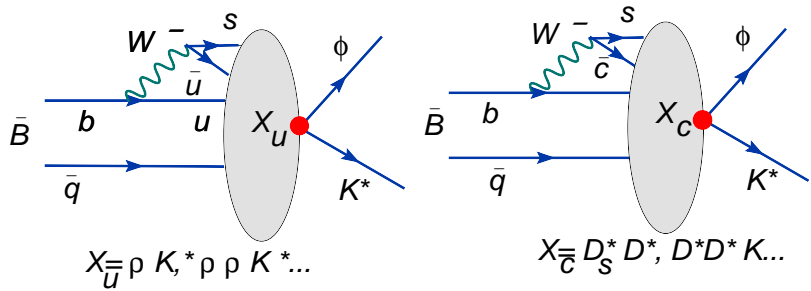}}
\end{tabular}
\caption{The rescattering  diagrams.} 
\label{fig:narrow2}
\end{center}
\end{figure}

Finally, we examine electroweak-penguin (EWP) contributions to
$B\to\phi K^*$. The standard EWP diagrams contribute mainly to $\fL$.
However, in Ref.~\cite{Beneke}, it was pointed out that
electromagnetic effects involving a photon that subsequently converts
to a vector meson can generate an unsuppressed transverse amplitude.
(Enhanced chromomagnetic dipole operators are discussed in
Ref.~\cite{HouMakiko}, with similar results as Ref.~\cite{Beneke}.)
With this new EWP mechanism, the observed value of $\fTfL$ in
$B\to\phi K^{*}$ may be explained, but it requires that this dipole
EWP contribution which enhances one of the transverse amplitudes by
$\alpha_{em}m_b / \Lambda_{\sss QCD}$ be sufficiently strong
\cite{Beneke}. Note that this electromagnetic contribution, and hence a
large value of $\fTfL$, should be observed in any decay where the
photon can convert into a neutral vector meson. However, a large
$\fTfL$ is observed in $B^+ \to \rho^+ K^{*0}$ decays (see
Table~\ref{table1}), but no EWP can contribute here. Thus, the new
enhanced EWP's cannot be the sole explanation of a large $\fTfL$. On
the other hand, in this paper we assume that there is a single
explanation for the large $\fTfL$'s, and so enhanced EWP contributions
of the nature discussed above are ruled out.

There are therefore only two proposed SM explanations of the observed
$\fTfL$ in $B\to\phi K^*$ decays: penguin annihilation and
rescattering. At this point, it is useful to make a general comment
about the two explanations. Penguin annihilation holds within a
specific calculation framework (QCDf). However, rescattering is just a
scenario -- there isn't even a concrete model. One can come up with a
particular model to implement rescattering \cite{soni}, but if it
fails, it doesn't rule out the idea -- one can simply invent other
models.

The naive expectation of small $\fTfL$ can be extended to the
hierarchy $|A_0|^2 \gg |A_+|^2 \gg |A_-|^2$. While both penguin
annihilation and rescattering ideas were proposed to explain the
violation of $|A_0|^2 \gg |A_\pm|^2$, the inequality $|A_+|^2 \gg
|A_-|^2$ may also be used to test the models. As we noted earlier, the
experimental observation of Table~\ref{table1} is indeed consistent
with $|A_+|^2 \gg |A_-|^2$. Simple models of rescattering \cite{soni}
violate this inequality, which is not supported by experimental
data. While this does not rule out the rescattering idea, this makes
it a less likely explanation. On the other hand, penguin annihilation
idea is consistent with $|A_+|^2 \gg |A_-|^2$.

Although the physical origin of penguin annihilation and rescattering
is different, the two explanations have similarities of calculation.
In order to see this, consider the penguin contribution ${\cal P}\!_q$
for the decay ${\bar b} \to {\bar q} q' {\bar q}'$ ($q=d,s$,
$q'=u,d,s$):
\begin{eqnarray}
\label{penguinamp}
{\cal P}\!_q & = & V_{ub}^* V_{uq} P_u + V_{cb}^* V_{cq} P_c
+ V_{tb}^* V_{tq} P_t \nn\\
& = & V_{cb}^* V_{cq} (P_c - P_u) + V_{tb}^* V_{tq} (P_t - P_u)~,
\end{eqnarray}
where the unitarity of the Cabibbo-Kobayashi-Maskawa (CKM) matrix has
been used in the second line.  In the rescattering solution the
dominant contribution to the transverse amplitudes come from $P_c$,
while the contributions from $P_{u,t}$ are small. In the penguin
annihilation solution the dominant contributions to the transverse
amplitudes come from $P_t$ through the penguin annihilation diagram,
and the contributions from $P_{u,c}$ are small. Thus, in either case,
the effect of the dominant contribution to the transverse amplitudes
is simply the addition of one amplitude. Below we follow this
prescription: we take into account the additional SM effects by adding
a single amplitude to represent the dominant contribution to the
transverse amplitudes.

\section{\boldmath $B \to \rho\rho$ Decays}

Both penguin annihilation and rescattering explain the $\fTfL$ ratio
in the $\btos$ decay $B\to\phi K^*$ by modifying the penguin
amplitude. A similar modification must appear in some $\btod$
decays. In this section we examine the predictions of these
explanations for $\fTfL$ in $B\to\rho\rho$ decays.  Experimental
measurements in $B\to\rho\rho$ along with related $B\to\rho\omega$,
$\omega\omega$, and $K^*\bar{K}^*$ decays are shown in Table~\ref{table2}. 
However, due to additional uncertainties, we do not consider modes 
with $\omega$ further in this paper. We will discuss $K^*\bar{K}^*$
decays in the next section.

\begin{table}[tbh]
\center
\begin{tabular}{lcccccc}
\hline
\vspace{-0.3cm}\\
Mode & ${\cal B}(10^{-6})$ & $f_{\sss L}$ & ~~ & $Tr$ & $C$ & $P$ \\
\vspace{-0.3cm}\\
\hline\hline
\vspace{-0.3cm}\\
$\rho^0\rho^+$ \cite{rho+rho0} &
  $18.2 \pm 3.0$  & ${0.912}^{+0.044}_{-0.045}$ & &
 $-1/\sqrt{2}$ & $-1/\sqrt{2}$ & $0$
 \\
$\rho^+\rho^-$ \cite{rho+rho-} &
  ${24.2}^{+3.0}_{-3.2}$ & $0.976^{+0.028}_{-0.024}$  & &
 $-1$ & $0$ & $-1$
 \\
$\rho^0\rho^0$ \cite{rho0rho0} &
  $(1.07 \pm 0.38)$  & $({0.86}^{+0.12}_{-0.14})$ & &
  $0$ & $-1/\sqrt{2}$ & $1/\sqrt{2}$
 \\
$\omega\rho^+$ \cite{omegakst} &
  ${10.6}^{+2.6}_{-2.3}$ & ${ 0.88}\pm{0.11}$  & &
 $1/\sqrt{2}$ & $1/\sqrt{2}$ & $\sqrt{2}$
 \\
$\omega\rho^0$ \cite{omegakst} &
  { $<1.5$} & --  & &
 $0$ & $0$ & $1$
 \\
$\omega\omega$ \cite{omegakst} &
  { $<4.0$} & -- & &
 $0$ & $1/\sqrt{2}$ & $1/\sqrt{2}$ 
 \\
$K^{*0}\bar{K}^{*0}$ \cite{kstkst} &
  { $<22$} & -- & &
 $0$ & $0$ & $1$
 \\
$K^{*+}\bar{K}^{*0}$ \cite{kstkst} &
  { $<71$} & --  & &
$0$ & $0$ & $1$
 \\
\vspace{-0.3cm}\\
\hline\hline
\end{tabular}
\caption{Measurements of the branching fraction ${\cal B}$ and
longitudinal polarization fraction $\fL$ for $B^+$ and $B^0_d$ meson
decays expected to proceed through the $\bar{b}\to\bar{d}$ transition
\cite{pdg, hfag}.  Numbers in parentheses indicate observables
measured with less than 4$\sigma$ significance.  The last three
columns show the naive amplitude decomposition in terms of
color-favored and color-suppressed tree amplitudes $Tr$ and $C$ and
the gluonic penguin amplitude $P$. }
\label{table2}
\end{table}

Within the diagrammatic approach \cite{GHLR}, the three $B \to
\rho\rho$ amplitudes are given mainly by three diagrams: the
color-favored and color-suppressed tree amplitudes $Tr$ and $C$, and
the gluonic penguin amplitude $P$.
\bea
-\sqrt{2} A(B^+ \to \rho^+ \rho^0) & = & Tr + C ~, \nn\\
- A(\bd \to \rho^+ \rho^-) & = & Tr + P ~, \nn\\
-\sqrt{2} A(\bd \to \rho^0 \rho^0) & = & C - P ~.
\eea
Since a modification of $P$ is involved, one sees immediately that
$\fTfL$ in $B^+ \to \rho^+ \rho^0$ will not be affected. This agrees
with observation (see Table~\ref{table2}).

As detailed in the previous section, the extra SM contribution is
taken into effect with the addition of a single amplitude, $R$:
\bea
- A(\bd \to \rho^+ \rho^-) & = & Tr + P + R ~, \nn\\
-\sqrt{2} A(\bd \to \rho^0 \rho^0) & = & C - P - R ~.
\eea
$Tr$, $C$, and $P$ contribute mainly to the $L$ polarization; the
contributions to $T$ arise at $O(1/m_b)$. In Ref~\cite{Beneke}, it was
pointed out that $C$ might contribute significantly to the transverse
amplitude through the hard-spectator scattering for certain choice of
the parameters representing this contribution. If one uses the default
values of these parameters in Ref~\cite{Beneke}, $C$ still contributes
dominantly to the $L$ polarization and so only $R$ contributes to
$T$. (The case where $C$ also contributes to $T$ is considered
below.) It is understood that any contributions to $T$ can be
different for $T=+,-$, so that there are two new contributions, $R_+$
and $R_-$. (As noted earlier, in $B \to \phi K^*$, $A_+ \gg A_-$. If
this were taken for $B \to \rho\rho$, we would have $R_+ \equiv R$,
$R_- \simeq 0$.)

In order to estimate $\fTfL$ for these decays, it is necessary to
estimate the size of $R_\pm$. As discussed earlier, rescattering
affects $P_c$, the $c$-quark contribution to the penguin
amplitude. Thus, $|R_\pm| \sim |P|$. (A similar conclusion holds for
$\btos$ decays). The measured value of $\fTfL$ in $B\to\phi K^*$ and
$B\to\rho K^*$ is explained if $|R'| \sim |P'|$.) For penguin
annihilation, the estimate is similar: $|R_\pm| \sim |P|$. Now, in
Ref.~\cite{GHLR}, the relative sizes of the $B \to \rho\rho$ diagrams
were roughly estimated as
\beq
1 : |Tr| ~~,~~~~ {\cal O}({\bar\lambda}) : |C|,~|P| ~,
\eeq
where ${\bar\lambda} \sim 0.2$. These estimates are expected to hold
approximately in the SM. This shows that $\fTfL$ is expected to be
small in $\bd \to \rho^+ \rho^-$, since it is proportional to
$(|R_+|^2 + |R_-|^2)/|Tr|^2 \sim |P|^2/|Tr|^2$. This agrees with
observation (see Table~\ref{table2}).

However, $\fTfL$ can be large in $\bd \to \rho^0 \rho^0$ since the
contributions to the transverse and longitudinal polarizations are the
same size. (However, see the estimate below.) It will be interesting
to measure this precisely.

There are further tests. Since there is only one added amplitude, one
has $|A_+(\bd \to \rho^0 \rho^0)| = |\Abar_-(\bdbar \to \rho^0
\rho^0)|$, and similarly for $A_-$ and $\Abar_+$. If this is not
found, penguin annihilation and rescattering will be ruled out.

We see that one can extract $|R_+|$ from $|A_+(\bd \to \rho^0
\rho^0)|$. As noted above, there are a variety of ways of obtaining
$|R'_+|$ from $\btos$ decays. One can then see if $|R_+|$ and $|R'_+|$ are
related by flavor SU(3), thus testing penguin annihilation and
rescattering. A similar exercise can be carried out for $|R_-|$ and
$|R'_-|$. Note that the effect of SU(3) breaking must be included in the
calculation.  While we do not know its size, it should not be very
large. A simple estimate of SU(3) breaking based on naive
factorization confirms this as all the vector mesons ($\rho$, $K^*$,
etc.) have masses and decay constants which are not very different.

In order to illustrate this, we use SU(3) to estimate $\fTfL$ in $\bd
\to \rho^0 \rho^0$ from $B^+ \to \rho^+ K^{*0}$ decays. The transverse
polarizations in these two modes are given by $R$ and $R'$,
respectively, where $R = \sqrt{|R_+|^2 + |R_-|^2}$, and similarly for
$R'$. $R$ and $R'$ are related by SU(3): with penguin annihilation, $R
= {\cal B} |V_{td}/V_{ts}| R'$, where ${\cal B}$ is the measure of
SU(3) breaking (with rescattering, the CKM ratio is $|V_{cd}/V_{cs}|$,
which is of the same order as $|V_{td}/V_{ts}|$). In what follows, we
neglect SU(3) breaking, so ${\cal B} = 1$. Now,
\bea
\fTfL (\bd \to \rho^0 \rho^0) & = & |A_{\sss T} (\bd \to \rho^0
\rho^0)|^2 / |A_{\sss L} (\bd \to \rho^0 \rho^0)|^2 \nn\\
& = & |V_{td}/V_{ts}|^2 |A_{\sss T} (B^+ \to \rho^+ K^{*0})|^2 /
|A_{\sss L} (\bd \to \rho^0 \rho^0)|^2 ~.
\eea
Using experimental data, we find
\bea
|A_{\sss T} (B^+ \to \rho^+ K^{*0})|^2 & = & (5.10 \pm 1.14) \times
 10^{-16}~{\rm GeV}^2 ~, \nn\\
|A_{\sss L} (\bd \to \rho^0 \rho^0)|^2 & = & (2.10 \pm 0.81) \times
 10^{-16}~{\rm GeV}^2 ~,
\eea
leading to
\beq
\fTfL (\bd \to \rho^0 \rho^0) = |V_{td}/V_{ts}|^2 \, (2.43 \pm 1.08) ~.
\eeq
There are two points to be made here. The first is that this agrees
with data taken directly from $\bd \to \rho^0 \rho^0$ (see
Table~\ref{table2}):
\beq
\fTfL (\bd \to \rho^0 \rho^0) = (1-\fL)/\fL = 0.16 \pm 0.15 ~.
\eeq
Because of the large errors, the agreement is good, showing that there
is no violation of SU(3). Equally, the measurement does not give a
definite answer as to whether $\fTfL$ is large or small. The second
point is related to this: if central values are taken, $\fTfL$ is not
large after all. This shows that $\fTfL$ is {\it not} guaranteed to be
large in $\bd \to \rho^0 \rho^0$. The reason for this is that, due to
the additional amplitude $C$, $\fL$ can be big, making $\fTfL$
small. If one wishes to ensure a large $\fTfL$, it is better to use
$\btod$ modes which are dominated by one amplitude in the SM: $P$.
This point will be used in the next section.

Finally, there is one more possible complication. Naively, the
contribution from the diagram $C$ to the transverse polarization is
suppressed by $O(1/m_b)$. However, as already
indicated earlier, in QCDf spectator corrections from the $C$ diagram,
which we denote as $C_{\sss T}$, may contribute significantly to the
transverse polarization \cite{Beneke}. If this is the case, there will
be two contributions to the transverse polarization: $R_i$ and
$C_{\sss T}^i$, $i=+,-$, with relative weak and strong
phases. Assuming that the weak phase is taken from independent
measurements, this leaves three parameters for a given transverse
polarization, say $T=+$. One thus needs three pieces of information in
$\bd \to \rho^0 \rho^0$ to obtain these parameters.

Unfortunately, at present, this is not possible. This can be
understood as follows. As above, $|A_+(\bd \to \rho^0 \rho^0)|$ and
$|\Abar_-(\bdbar \to \rho^0 \rho^0)|$ provide two measurements. The
third piece of information would be to find the relative phase of
these two amplitudes. Now, the angular analysis of $B\to V_1V_2$
decays allows one to extract ${\rm Im}(A_\perp A_0^*)$, ${\rm
Im}(A_\perp A_\|^*)$, and ${\rm Re}(A_0 A_\|^*)$ \cite{BVVTP}. This
gives the relative phases of the $A$ amplitudes. A similar exercise
can be carried out for ${\bar B}\to {\bar V}_1{\bar V}_2$, giving the
relative phases of the ${\bar A}$ amplitudes. Note that this does not
give the relative phases of the $A$ and ${\bar A}$ amplitudes.
However, for $\btos$ decays, $A_0 = {\bar A}_0$. Thus, the angular
analysis of both $\bd \to \phi K^*$ and $\bdbar \to \phi {\bar K}^*$
{\it does} allow one to obtain the relative strong phases of all
amplitudes. (This has been carried out, and is how $\phi_\perp$ and
$\phi_\parallel$ of Table~\ref{table1} were obtained in $B\to\phi
K^{*}$.) But the same technique cannot be used for $\btod$ decays,
whose longitudinal polarization involves two amplitudes
[Eq.~(\ref{penguinamp})]. In order to obtain the relative strong
phases of $A$ and ${\bar A}$ amplitudes in $\btod$ decays, it will be
necessary to perform a time-dependent angular analysis (this is
described in detail in Sec.~5). This is possible, but it is a future
measurement.

We therefore conclude that it is extremely difficult to perform the
tests of penguin annihilation and rescattering described above if
$C_{\sss T}^i$ contributions are present. The lesson here is that it
is best to consider $\btod$ decays for which $\fTfL$ is expected to be
large and which receive only one dominant contribution to the
transverse polarization. We will return to this point in the next
section.

\section{U-Spin Pairs}

In the past sections, we have stressed the idea of measuring $\fTfL$
in $\btod$ decays. But this raises the question: how does one choose
the $\btod$ decay to study? One tool which is very useful in this
regard is U-spin. U-spin is the symmetry that places $d$ and $s$
quarks on an equal footing, and is often given as transposing $d$ and
$s$ quarks: $d \leftrightarrow s$. Pairs of $B$ decays which are
related by U-spin are given in Ref.~\cite{Gronau}. In $B\to VV$ form,
these are
\begin{enumerate}
\item $\bd \to K^{*+}\rho^-$ and $\bs \to \rho^+ K^{*-}$~,
\item $\bs \to K^{*+} K^{*-}$ and $\bd \to \rho^+\rho^-$~,
\item $\bd \to K^{*0}\rho^0$ and $\bs \to \Kbar^{*0}\rho^0$~,
\item $B^+ \to K^{*0} \rho^+$ and $B^+ \to \Kbar^{*0} K^{*+}$~,
\item $\bs \to K^{*0} \Kbar^{*0}$ and $\bd \to \Kbar^{*0} K^{*0}$~.
\end{enumerate}
In all cases, the first decay is $\Delta S = 1$ ($\btos$); the second
is $\Delta S = 0$ ($\btod$). Annihilation-type decays have been
ignored. The procedure here is straightforward: one must measure the
polarizations in the $\btos$ decay, and compare them with the
measurements in the corresponding $\btod$ decay.

As noted previously, the best $\btod$ decays are those for which
$\fTfL$ is expected to be large and which receive only one dominant
contribution to the transverse polarization. Keeping only the largest
contributions, the SM amplitudes for the $\Delta S = 0$ decays are
\begin{enumerate}
\item $A(\bs \to \rho^+ K^{*-}) = -\left[Tr + P \right]$~,
\item $A(\bd \to \rho^+\rho^-) = -\left[Tr + P \right]$~,
\item $\sqrt{2} A(\bs \to \Kbar^{*0}\rho^0) = -\left[C-P\right] $~,
\item $A(B^+ \to \Kbar^{*0} K^{*+}) = P$~,
\item $A(\bd \to \Kbar^{*0} K^{*0}) = P$~.
\end{enumerate}
The (potential) significant contributions to the transverse
polarization $+$ are
\begin{enumerate}
\item $\bs \to \rho^+ K^{*-}: Tr_{\sss T}^+, R_+$ ~,
\item $\bd \to \rho^+\rho^-: Tr_{\sss T}^+, R_+$~,
\item $\bs \to \Kbar^{*0}\rho^0: C_{\sss T}^+, R_+$~,
\item $B^+ \to \Kbar^{*0} K^{*+}: R_+$~,
\item $\bd \to \Kbar^{*0} K^{*0}: R_+$~,
\end{enumerate}
and similarly for $T=-$.

The first two decays are dominated by the tree diagram and are
therefore expected to show small transverse polarization. This is
reflected in the polarization measurement of $\bd \to \rho^+\rho^-$ (
Table~\ref{table2}). The remaining three decays can have a large
$\fTfL$.

In the past section, we have argued that it is best to consider
$\btod$ decays which receive only one dominant contribution to the
transverse polarization, i.e.\ they are dominated by $P$ in the
SM. Given this, the best possibilities are the last two. We therefore
consider (i) $B^+ \to K^{*0} \rho^+$ ($\btos$) and $B^+ \to \Kbar^{*0}
K^{*+}$ ($\btod$) and (ii) $\bs \to K^{*0} \Kbar^{*0}$ ($\btos$) and
$\bd \to \Kbar^{*0} K^{*0}$ ($\btod$). We urge the measurement of
$\fTfL$ in these pairs of decays.

The explanations of $\fTfL$ in $B\to\phi K^*$ then make three
predictions :
\begin{itemize}

\item $\fTfL$ is expected to be large in both the $\btos$ decay and
the corresponding $\btod$ decay.

\item $|A_{+}|$ and $|{\bar A}_{-}|$ are expected to be equal in both the
$B$ and ${\bar B}$ decays, and similarly for $A_{-}$ and
${\bar A}_{+}$.

\item $R'_i$ and $R_i$ ($i=+,-$) can be extracted from the $\btos$ and
$\btod$ decays, respectively. These should be related by flavor SU(3)
(including SU(3) breaking).

\end{itemize}
If any of these predictions fail, penguin annihilation and
rescattering are ruled out in the U-spin limit or for small U-spin
breaking.

The ratio of $\fTfL$ in these pairs of decays measures SU(3)
breaking. For a given transverse polarization,
\beq
({\fTfL})_{\btod}/({\fTfL})_{\btos} = (|R_i|^2/|R'_i|^2)/(|P_{\sss L}|^2/|P'_{\sss L}|^2) ~,
\eeq
where $P_{\sss L}$ and $P'_{\sss L}$ are the longitudinal parts of the
penguin diagram in $\btod$ and $\btos$ transitions. Although one
cannot prove it rigorously, it is likely that the SU(3) breaking in
$(|R_i|/|R'_i|)$ is of the same size as that in $(|P_{\sss L}|/|P'_{\sss
L}|)$, so that the net SU(3) breaking in this ratio is small. If one
ignores SU(3) breaking for this reason, another prediction which can
be used to test penguin annihilation and rescattering is that
\beq
({\fTfL})_{\btod} = ({\fTfL})_{\btos} ~.
\label{test}
\eeq
The breaking of SU(3) in the above equation is model dependent and, as
indicated above, we expect to find small SU(3) breaking in
Eq.~\ref{test} in models of penguin annihilation or rescattering. If it
is found experimentally that the above relation is broken badly, then
these models will have to invent a mechanism to generate large
SU(3)-breaking effects or they will be ruled out. In other words,
Eq.~\ref{test} can be used to constrain specific models of penguin
annihilation and rescattering.

\section{Distinguishing Penguin Annihilation and \\ Rescattering}

Up to now, we have not distinguished penguin annihilation and
rescattering, arguing that their effects are very similar. However, is
it possible to differentiate these two scenarios? As we will see in
the present section, the answer is yes.

As noted above, rescattering involves only a change to $P_c$, while
penguin annihilation involves only $P_t$. However, the weak phase of
these pieces is different: $\phi$(rescattering)$\sim 0$,
$\phi$(penguin annihilation)$\sim -\beta$, If this weak phase can be
measured, one can distinguish penguin annihilation and rescattering.

This can be done as follows. Consider a penguin-dominated $\btod$
decay in which the transverse polarization is observed to be large.
Within the SM, this would be the result of a single dominant
contribution originating from large rescattering or penguin
annihilation.  If the transverse amplitude in the penguin-dominated
$\btod$ decay is small then either rescattering or penguin annihilation
is ruled out, or there is large SU(3) breaking in Eq.~\ref{test}.
Regardless, for small measured transverse polarization, we cannot
assume the transverse amplitude to be dominated by a single
contribution and so henceforth we will assume that a large transverse
amplitude is observed in the penguin-dominated $\btod$ decay. The
transverse amplitude is then dominated by a single amplitude and we
can parameterize this contribution as $R e^{i\phi} e^{i\delta}$, where
$\phi$ and $\delta$ are the weak and strong phases, respectively. The
transverse-polarization contribution to the CP-conjugate decay is then
$R e^{-i\phi} e^{i\delta}$. Thus, the ratio of the
transverse-polarization amplitude in the $\btod$ and CP-conjugate
decays is $e^{-2i\phi}$. In other words, this ratio measures the weak
phase and allows us to distinguish penguin annihilation and
rescattering.

In order to obtain this information, one needs to measure the relative
phase of $A_{\sss T}(\btod~{\rm decay})$ and $\Abar_{\sss T}( b \to d
~{\rm decay})$. As discussed earlier, this can be obtained by
performing a time-dependent angular analysis of the $\btod$ decay. We
give details of the procedure below.

Using CPT invariance, the full decay amplitudes for $B \to V_1V_2$ can
be written as \cite{glambda, DLNP2}
\bea
{\cal A} &=& Amp (\bvv) = A_0 g_0 + A_\| g_\| + i \, A_\perp
g_\perp~, \nn\\
{\bar{\cal A}} &=& Amp ({\bar B} \to V_1 V_2) = {\bar A}_0 g_0 + {\bar
A}_\| g_\| - i \, {\bar A}_\perp g_\perp~,
\label{fullamps}
\eea
where the $g_\lambda$ are the coefficients of the helicity amplitudes
written in the linear polarization basis. The $g_\lambda$ depend only
on the angles describing the kinematics \cite{glambda}. 

Using the above equations, we can write the time-dependent decay rates
as
\beq
\Gamma(\bbarp(t) \to V_1V_2) = e^{-\Gamma t} \sum_{\lambda\leq\sigma}
\Bigl(\Lambda_{\lambda\sigma} \pm \Sigma_{\lambda\sigma}\cos(\Delta M
t) \mp \rho_{\lambda\sigma}\sin(\Delta M t)\Bigr) g_\lambda g_\sigma
~.
\label{decayrates}
\eeq
Thus, by performing a time-dependent angular analysis of the decay
$B(t) \to V_1V_2$, one can measure 18 observables (not all are
independent). These are:
\bea
\Lambda_{\lambda\lambda}=\displaystyle
\frac{1}{2}(|A_\lambda|^2+|{\bar A}_\lambda|^2),~~&&
\Sigma_{\lambda\lambda}=\displaystyle
\frac{1}{2}(|A_\lambda|^2-|{\bar A}_\lambda|^2),\nn \\[1.ex]
\Lambda_{\perp i}= -\!{\rm Im}({ A}_\perp { A}_i^* \!-\! {\bar
A}_\perp {{\bar A}_i}^* ),
&&\Lambda_{\| 0}= {\rm Re}(A_\| A_0^*\! +\! {\bar A}_\| {{\bar A}_0}^*
), \nn \\[1.ex]
\Sigma_{\perp i}= -\!{\rm Im}(A_\perp A_i^*\! +\! {\bar A}_\perp
{{\bar A}_i}^* ),
&&\Sigma_{\| 0}= {\rm Re}(A_\| A_0^*\!-\! {\bar A}_\| {{\bar A}_0}^*
),\nn\\[1.ex]
\rho_{\perp i}\!=\! {\rm Re}\!\Bigl(e^{-i\phi^q_{\sss M}} \!\bigl[A_\perp^*
{\bar A}_i\! +\! A_i^* {\bar A}_\perp\bigr]\Bigr),
&&\rho_{\perp \perp}\!=\! {\rm Im}\Bigl(e^{-i\phi^q_{\sss M}}\, A_\perp^*
{\bar A}_\perp\Bigr),\nn\\[1.ex]
\rho_{\| 0}\!=\! -{\rm Im}\!\Bigl(e^{-i\phi^q_{\sss M}}[A_\|^* {\bar A}_0\! +
\!A_0^* {\bar A}_\| ]\Bigr),
&&\rho_{ii}\!=\! -{\rm Im}\!\Bigl(e^{-i\phi^q_{\sss M}} A_i^* {\bar
A}_i\Bigr),
  \label{eq:obs}
\eea
where $i=\{0,\|\}$ and $\phi^q_{\sss M}$ is the weak phase factor
associated with $B_q^0$--${\bar B}_q^0$ mixing. Note that the signs of
the various $\rho_{\lambda\lambda}$ terms depend on the CP-parity of
the various helicity states. We have chosen the sign of $\rho_{ii}$ to
be $-1$, which corresponds to the final state $\phi K^*$. The
quantities $\rho_{a,a}$, where $a= \|, \perp$, are sensitive to the
weak phase between $A_{\sss T}$ and $\Abar_{\sss T}$. Hence, for the
case of penguin annihilation, the quantities $\rho_{a,a} /
\Lambda_{a,a}$ are zero, while for rescattering these quantities are
nonzero and equal $ \pm\sin{2 \beta}$. (Note that since $A_{\sss T}$ is
dominated by a single amplitude, we have $|A_a|= |\bar{A}_a|$.)

Which $\btod$ decay should be used? It is necessary to consider one
which receives only one dominant contribution to the transverse
polarization and for which a time-dependent angular analysis can be
done. Of the decays studied in the previous sections, there is only
one which satisfies these requirements: $\bd \to \Kbar^{*0} K^{*0}$.
Thus, the measurement of the time-dependent angular analysis here
would allow one to distinguish penguin annihilation and rescattering,
and we urge experimentalists to look at this.  This time-dependent
angular analysis can be performed with the all-charged-track final
state $\bd \to \Kbar^{*0} K^{*0} \to K^-\pi^+K^+\pi^-$ without the
need to reconstruct $K^{*0}\to K_S^0\pi^0$ decays, which should
facilitate experimental measurements once this decay is observed.

Finally, we note that even without the time-dependent analysis, the
full angular analysis of $B \to \Kbar^{*} K^{*}$ decays could help in
distinguishing the models. If the strong phase difference between the
$(P_c - P_u)$ and $(P_t - P_u)$ amplitudes in Eq.~\ref{penguinamp} is
small, then the $\Delta\phi_{\perp,\parallel}$ parameters~\cite{pdg,
phiK*0} will be either negative or positive depending on the penguin
annihilation or rescattering model. However, if the strong phase is
not small, this will result in large direct CP violation in the $A_0$
amplitude which would be measured. A large strong phase would make it
difficult to resolve ambiguities in $\Delta\phi_{\perp,\parallel}$,
but limits on direct CP violation could constrain the strong phase
difference. We also note that a similar method in $\bar{b}\to\bar{s}$
decays, such as $B\to\phi K^*$, does not work because both penguin
annihilation and rescattering result in
$\Delta\phi_{\perp,\parallel}=0$, consistent with present data
\cite{phiK*0,belle:phikst,babar:phikstpl}.

\section{Conclusions}

The final-state particles in $B\to\phi K^*$ are vector mesons ($V$),
which means that this decay is in fact three separate decays, one for
each polarization of the $V$ (one longitudinal, two transverse).
Naively, it is expected that the fraction of transverse decays, $\fT$,
is much less than the fraction of longitudinal decays, $\fL$. However,
it is observed that these fractions are roughly equal: $\fTfL \simeq
1$. This is the $B\to\phi K^*$ polarization puzzle.

Other, similar, polarization puzzles have been measured, all in
$\btos$ decays. Within the standard model, there have been several
explanations of these results. However, if one requires a single
explanation of all polarization puzzles, two possibilities remain:
penguin annihilation and rescattering. Both of these also predict
large $\fTfL$ in certain $\btod$ decays. Indeed, by looking at $\btod$
decays, it is possible to test penguin annihilation and rescattering.
This is the purpose of this paper.

We begin with $B \to \rho\rho$ decays. We show that $\fTfL$ is
expected to be small in $B^+ \to \rho^+ \rho^0$ and $\bd \to \rho^+
\rho^-$; it is only in $\bd \to \rho^0\rho^0$ that $\fTfL$ can be
large, though this is not guaranteed. Although penguin annihilation
and rescattering are physically different, mathematically they are
similar. If it is found that $\fTfL$ is large in $\bd \to
\rho^0\rho^0$, it may be possible to test these explanations. For
example, if one compares penguin annihilation or rescattering in $\bd
\to \rho^0\rho^0$ with that found in $\btos$ decays, one can see if
flavor SU(3) is respected.

Because large effects are not ensured in $\bd \to \rho^0\rho^0$, it is
useful to consider other $\btod$ decays. We examine those which are
related by U-spin to other $\btos$ decays. We find two promising
U-spin pairs: (i) $B^+ \to K^{*0} \rho^+$ ($\btos$) and $B^+ \to
\Kbar^{*0} K^{*+}$ ($\btod$) and (ii) $\bs \to K^{*0} \Kbar^{*0}$
($\btos$) and $\bd \to \Kbar^{*0} K^{*0}$ ($\btod$). A large $\fTfL$
is predicted by penguin annihilation or rescattering in these
decays. In addition, the measurement of $\fTfL$ in these pairs of
decays will allow us to test these explanations by seeing if flavor
SU(3) is respected.

Up to now, we have treated penguin annihilation and rescattering as
similar. However, it is possible to distinguish penguin annihilation
from rescattering by performing a time-dependent angular analysis of
$\bd \to \Kbar^{*0} K^{*0}$. This is difficult experimentally, but it
may be possible at a future machine.

\bigskip
\noindent
{\bf Acknowledgments}:
This work was financially supported by NSERC of Canada (DL, MN \& AS),
and the U.S. NSF and A.~P.~Sloan Foundation (AG).


\end{document}